\documentclass[%
 reprint,
 amsmath,amssymb,
 aps,
]{revtex4-1}

\usepackage{graphicx}
\usepackage{dcolumn}
\usepackage{bm}
\usepackage{subfigure}
\usepackage{url}
\usepackage{float}

\begin{document}

\preprint{APS/123-QED}

\title{Vortex Dynamics in Type II Superconductors}

\author{Dachuan Lu}
\affiliation{%
 (Kuang Yaming Honors School, Nanjing University)
}%


\date{\today}

\begin{abstract}
Time dependent Ginzburg-Landau equation is solved for type II superconductors numerically, and the dynamics of entering vortices, geometric defects and pinning effects have been investigated. A superconducting wire with ratchet defects is designed to pump the vortices move in a specific direction and enhance the supercurrent when applying the periodical magnetic field. Some properties of this wire have been investigated numerically and analytically.
\end{abstract}

\maketitle


\section{Introduction}
High temperature superconductors(HTSC) are widely used in industrial production and medical equipments. Almost all the HTSCs are type II superconductors, the vortex pinning effects and vortex dynamic need concerning\cite{blatter1994vortices}.\par
Ginzburg and Landau proposed a phenomenological theory for superconducting phase\cite{landau1950theory} based on Landau second-order phase transitions, and the so-called Ginzburg-Landau equation is widely used to study the vortex in type II superconductors after Abrikosov firstly predicted the vortex lattice in type II superconductors \cite{blatter1994vortices}\cite{abrikosov1952influence}. Geometry and pinning has large effects on the configuration of vortex lattice. \par
Time dependent Ginzburg-Landau (TDGL) equation has more fascinating properties than the original stationary one\cite{schmid1966time}, and can be used to investigate the dynamic properties of the vortices. Mu\cite{mu1997linearized} and Du\cite{du2005numerical} have solved the TDGL equation by finite element method, Pedersen use COMSOL(finite element solver) to solve the TDGL in complicated geometry\cite{alstrom2011magnetic}. Some other methods are widely used to investigate the vortex dynamics. Reichhardt has used molecular dynamics(MD) to simulate the microscopic behavior of flux and calculate some macroscopic quantities, such as magnetization hysteresis and critical current density\cite{reichhardt1996vortex} \cite{richardson1994confirmation} \cite{reichhardt1998commensurate}. They modeled the vortex-vortex and vortex-pin force and calculated the total force exerted to each vortex. However, Ginzburg-Landau theory is more suitable for the situation with varying magnetic field, because more vortices will penetrate into the sample while increasing the magnetic field, and molecular dynamics method could hardly calculate the creation and annihilation of the vortices.\par
In Section.~\ref{tdgl}, we normalize the TDGL to the simplest form and In Section.~\ref{NS}, we present some numerical results based on the normalized TDGL, the dynamics of entering vortices, geometric defects and pinning effects have been investigated. In Section.~\ref{fp}, we investigate the flux pump which is an application based on our former investigation and can pump the vortices to move in specific direction. In Section.~\ref{sum}, we give a summary of this paper and clarify some limitation of our work.

\section{Time Dependent Ginzburg-Landau Equation}\label{tdgl}
In 1950, Ginzburg and Landau proposed the famous phenomenological theory\cite{landau1950theory} to describe the superconductors, in which, complex order parameter is introduced to describe the Cooper pair and normal electrons, a $\phi-4$ like potential is considered which accounts for the phase transition. Ginzburg-Landau theory can precisely describe the magnetic field penetration through the quantized magnetic flux vortices formation in the type II superconductors.\par

To find the dynamics of the vortex in superconductors, we turn to the time dependent Ginzburg Landau (TDGL) equation which is generalized by Schmid\cite{schmid1966time}. The TDGL equation is read as,
\begin{widetext}
\begin{equation}\label{tdgl1}
\frac{\hbar }{{2mD}}\left( {\frac{\partial }{{\partial t}} + i\frac{q}{\hbar }\Phi } \right)\Psi  =  - \frac{1}{{2m}}{\left( {\frac{\hbar }{i}\nabla  - q{\bf{A}}} \right)^2}\Psi  + \alpha \Psi  - \beta {\left| \Psi  \right|^2}\Psi
\end{equation}
\begin{equation}\label{tdgl2}
\sigma \left( {\frac{{\partial {\bf{A}}}}{{\partial t}} + \nabla \Phi } \right) = \frac{{q\hbar }}{{2mi}}\left( {{\Psi ^*}\nabla \Psi  - \Psi \nabla {\Psi ^*}} \right) - \frac{{{q^2}}}{m}{\left| \Psi  \right|^2}{\bf{A}} - \frac{1}{{{\mu _0}}}\nabla  \times \left( {\nabla  \times {\bf{A}} - {{\bf{B}}_{\bf{a}}}} \right)
\end{equation}
\end{widetext}
where $\Psi$ is the order parameter, $\mathbf{A}$ and $\Phi$ are the vector potential and scalar potential of the electromagnetical field, $\mathbf{B_a}$ is the applied magnetic field, $m$ and $q$ are the mass and charge of the Cooper pair respectively, and $\sigma$ is the conductance of the normal electrons. $\alpha, \beta, D$ are the parameter dependent on the properties of superconductors, in which $\beta$ is assumed to be a constant while $\alpha(T)=\alpha(0)(1-\frac{T}{T_c})$ is dependent on the temperature $T$.\par
The LHS(left hand side) of Eq.~\ref{tdgl1} and Eq.~\ref{tdgl2} are time dependent term, and the RHS are original Ginzburg-Landau theory. To describe the superconductor in vacuum, we choose the following appropriate boundary conditions,
\begin{equation}\label{bc1}
\left\{ \begin{array}{l}
\left( {\frac{\hbar }{i}\nabla \Psi  - q{\bf{A}}\Psi } \right) \cdot {\bf{n}} = 0\\
\nabla  \times {\bf{A}} = {{\bf{B}}_{\bf{a}}}\\
\left( {\frac{{\partial {\bf{A}}}}{{\partial t}} + \nabla \Phi } \right) \cdot {\bf{n}} = 0
\end{array} \right.
\end{equation}
The first and the last equation in Eq.~\ref{bc1} implies that the supercurrent and the normal current is zero at the boundary, and the second equation illustrates that the magnetic induction is equal to the applied magnetic field at the boundary.
\subsection{Normalization}
For simplicity in simulation, the parameters in Eq.~\ref{tdgl1} and Eq.~\ref{tdgl2} can be arranged into dimensionless quantities by introducing London penetration depth $\lambda$, Ginzburg-Landau coherence length $\xi$ and Ginzburg-Landau parameter $\kappa=\lambda/\xi$. The corresponding quantities are shown as follows,
\begin{equation}\label{dimless}
\begin{array}{l}
\left( {x,y,z} \right) \to \left( {\lambda x',\lambda y',\lambda z'} \right),t \to \frac{{{\xi ^2}}}{D}t'\\
A = \frac{\hbar }{{q\xi }}A',\Phi  = \alpha D{\kappa ^2}\sqrt {\frac{{2{\mu _0}}}{b}} \Phi ',\\
\Psi  = \sqrt {\frac{\alpha }{\beta }} \Psi ',\sigma  = \frac{1}{{{\mu _0}D{\kappa ^2}}}\sigma '
\end{array}
\end{equation}
By applying the normalization procedure, the TDGL equations can be read as,
\begin{widetext}
\begin{equation}\label{ngl1}
\left( {\frac{\partial }{{\partial t}} + i\kappa \Phi } \right)\Psi  =  - {\left( {\frac{i}{\kappa }\nabla  + {\bf{A}}} \right)^2}\Psi  + \Psi  - {\left| \Psi  \right|^2}\Psi
\end{equation}
\begin{equation}\label{ngl2}
\sigma \left( {\frac{{\partial {\bf{A}}}}{{\partial t}} + \nabla \Phi } \right) = \frac{1}{{2i\kappa }}\left( {{\Psi ^*}\nabla \Psi  - \Psi \nabla {\Psi ^*}} \right) - {\left| \Psi  \right|^2}{\bf{A}} - \nabla  \times \left( {\nabla  \times {\bf{A}} - {{\bf{B}}_{\bf{a}}}} \right)
\end{equation}
\end{widetext}
and the boundary conditions are,
\begin{equation}\label{nbc}
\begin{array}{l}
\left( {\frac{\hbar }{i}\nabla \Psi  + {\bf{A}}\Psi } \right) \cdot {\bf{n}} = 0, \\
\nabla  \times {\bf{A}} = {{\bf{B}}_{\bf{a}}}, \\
\left( {\frac{{\partial {\bf{A}}}}{{\partial t}} + \nabla \Phi } \right) \cdot {\bf{n}} = 0,
\end{array}
\end{equation}
\subsection{Gauge Invariance}
Because the free energy is invariant under the $U(1)$ gauge transformation,
\begin{equation}\label{gaugetrans}
\Psi ' = \Psi {e^{i\kappa \chi }},A' = A + \nabla \chi ,\Phi ' = \Phi  - \frac{{\partial \chi }}{{\partial t}}
\end{equation}
where $\chi$ is an arbitrary function, we can fix the gauge by varying the $\chi$ and make the scalar potential $\Phi$ to be zero, namely,
\[\frac{{\partial \chi }}{{\partial t}} = \Phi. \]
The terms containing $\Phi$ in the Ginzburg-Landau equation(Eq.~\ref{ngl1}, Eq.~\ref{ngl2}) and the boundary conditions (Eq.~\ref{nbc}) can be dropped.\par
The $U(1)$ gauge invariance implies that the particle number is conserved in this system, for the Hamiltonian is commuted with the particle number operator $\hat{Q}=\mathop \sum \limits_n {\Psi _n}^\dag {\Psi _n}$. This property indicates that there is no dissipation in this system.

\section{Numerical Simulation}\label{NS}
In this section, we investigate the numerical solution of dimensionless TDGL(Eq.~\ref{ngl1}, Eq.~\ref{ngl2}) by using the finite element method which is implemented by COMSOL \cite{zimmerman2006multiphysics}. For simplicity but with universality, we consider 2D geometry because the superconducting layers of many HTSCs have strong 2D properties. We set the superconducting layer in $xy$ plane and designate $(x,y,t)$ as the spatial and temporal coordinate. The magnetic field is set along the $z$ direction.\par
In numerical implementation, the order parameter is divided into real part and imaginary part, $\Psi=\Re{\Psi}+i \Im{\Psi}$, the vector potential is decomposed into scalar functions in different directions, namely $\mathbf{A}=A_x \mathbf{i}+A_y \mathbf{j}$.\par
In the following section, we will first investigate some universal properties of the vortex dynamics, and then turn to the flux pump, which is the application of the vortex dynamics and intends to enhance the supercurrent.

\subsection{Disk Sample with SO(2) symmetry }
\begin{figure}[H]\label{k1xf}
\centering
\includegraphics[width=0.4\textwidth]{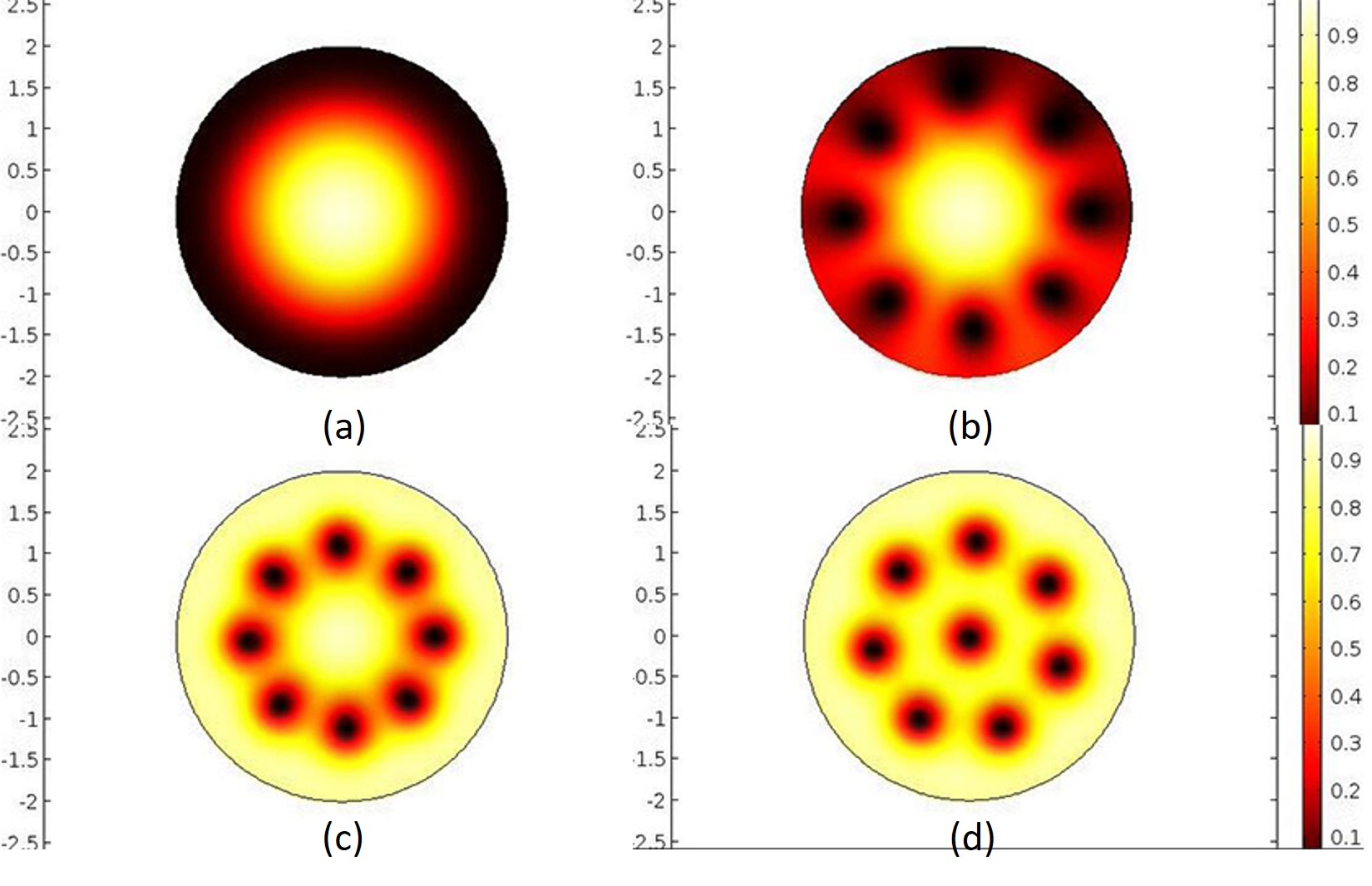}
\caption{The evolution of vortices penetrating into the disk sample. we set $\kappa=4, B_a=1.5, \sigma=1$ and the diameter is $d=4$, they are all dimensionless parameter. The density plot illustrates the magnitude of order parameter, the higher magnitude, the lighter color, (a) the magnetic field penetrates into the sample uniformly and destroy the Meissner state, (b) the magnetic field tends to exist in form of flux vortices, (c) the vortices get into the sample, (d) the vortices are rearranged to a low energy configuration.}
\label{cirf}
\end{figure}

The disk remains unchanged under the rotation of any angle, and when we apply the magnetic field along the $z$-axis as described in the Section.~\ref{NS}, the vortices will form from the boundary and get into the sample, consistent with the some previous works\cite{schweigert1999flux}\cite{baelus2001saddle}. Because of the system tends to the minimum free energy, the vortices will be rearranged to reduce the interaction energy, but the original symmetry is broken. The evolution process of the vortices is shown in Fig.~\ref{cirf}.\par
\begin{figure}[H]
\centering
\includegraphics[width=0.40\textwidth]{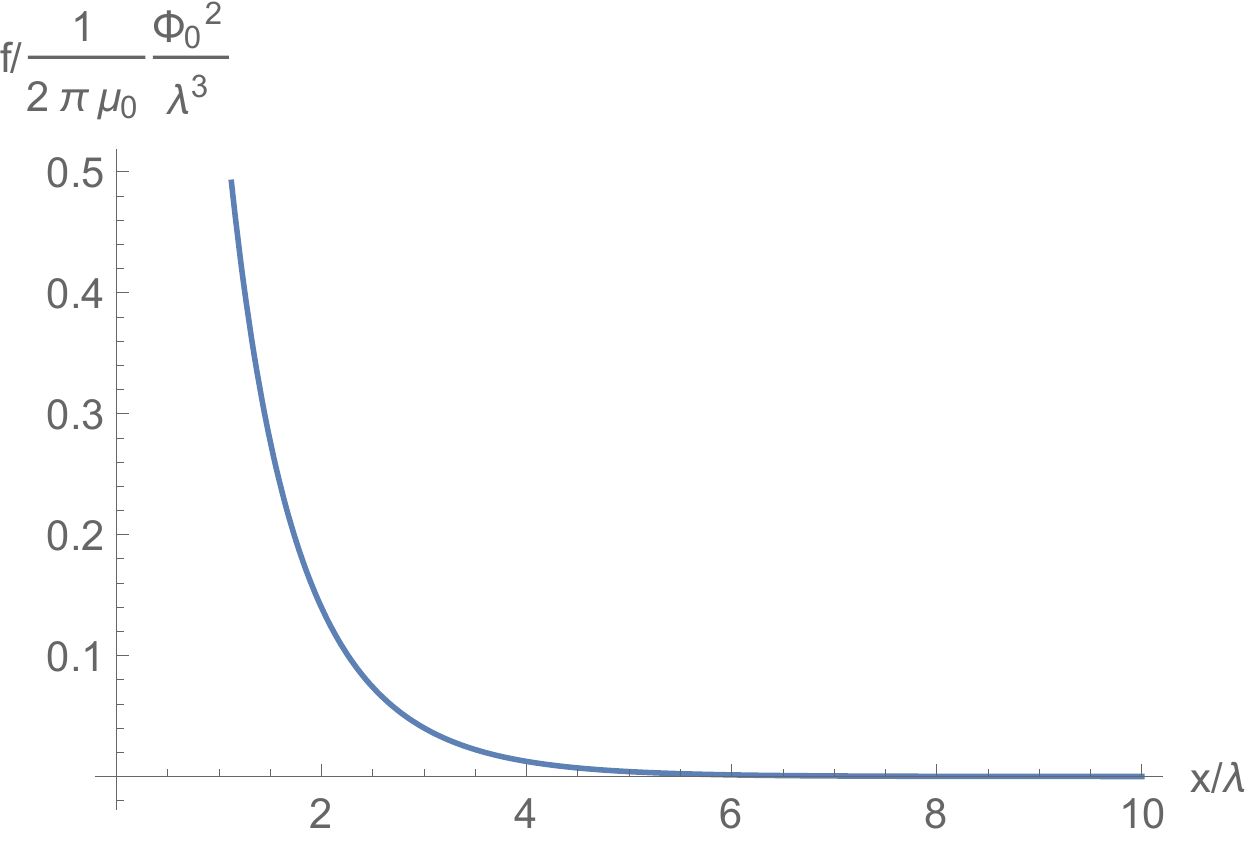}
\caption{Repulsive force of the adjacent vortices }
\label{k1x}
\end{figure}

The vortices tends to repel each other, and the repulsive force is $\frac{1}{{2\pi {\mu _0}}}\frac{{{\Phi _0}^2}}{{{\lambda ^3}}}{K_1}\left( {\frac{x}{\lambda }} \right)$, which decays with the distance $x$ dramatically, as Fig.~\ref{k1x} illustrates. The first order modified bessel function of the second kind, ${{\rm{K}}_1}\left( {\frac{r}{\lambda }} \right) \approx \frac{\lambda }{r}$ as $r \rightarrow 0$ and ${{\rm{K}}_1}\left( {\frac{r}{\lambda }} \right) \approx \sqrt {\frac{{\pi \lambda }}{{2r}}} {e^{ - r/\lambda }}$ as $r\rightarrow \infty$. The corresponding short range interaction is large, while the long range interaction decays fast, so we could set a cutoff in numerical simulation.\par

We also calculate the energy density of this process by using the following equations,
\begin{equation}
\begin{array}{l}
{H_{\sup }} = \frac{1}{{{\kappa ^2}}}{\left| {\nabla \Psi } \right|^2} - {\left| \Psi  \right|^2} + \frac{1}{2}{\left| \Psi  \right|^4}\\
{H_{{\rm{mag}}}} = {\left( {{{\bf{B}}_a} - \nabla  \times {\bf{A}}} \right)^2}\\
{H_{{\mathop{\rm int}} }} = \frac{i}{\kappa }{\bf{A}}\left( {\left( {\nabla \Psi } \right){\Psi ^*} - \Psi \left( {\nabla {\Psi ^*}} \right)} \right) + {\left| {\bf{A}} \right|^2}{\left| \Psi  \right|^2}
\end{array}
\end{equation}
where ${H_{\sup }}$ is the superconducting energy, ${H_{{\rm{mag}}}}$ is the magnetic energy and ${H_{{\mathop{\rm int}} }}$ is the interaction energy. The total energy is,
\begin{equation}
{H_{{\rm{tot}}}} = {H_{\sup }} + {H_{{\rm{mag}}}} + {H_{{\mathop{\rm int}} }}
\end{equation}

\begin{figure}[H]
\centering
\includegraphics[width=0.4\textwidth]{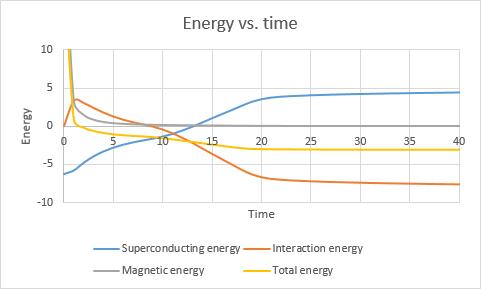}
\caption{Different parts of energy change along with time. The blue line is the superconducting energy, increases dramatically when the vortices are entering the sample. The red line is the interaction energy and has a maximum at the point when the vortices are entering the sample. The grey line is the magnetic energy and decreases along with the magnetic field penetrating into the sample and forming vortices. The yellow line is the total energy and always decreases to find the lowest energy state.}
\label{energytime}
\end{figure}

\subsection{Disk Sample with Defect}\label{defecteffect}
From the previous section, we can gain some general idea of how vortices form and evolve, in this section, we will investigate the sample with geometrical defects which will break the symmetry and make the motion of vortices anisotropy. \par
\begin{figure}[H]
\centering
\includegraphics[width=0.4\textwidth]{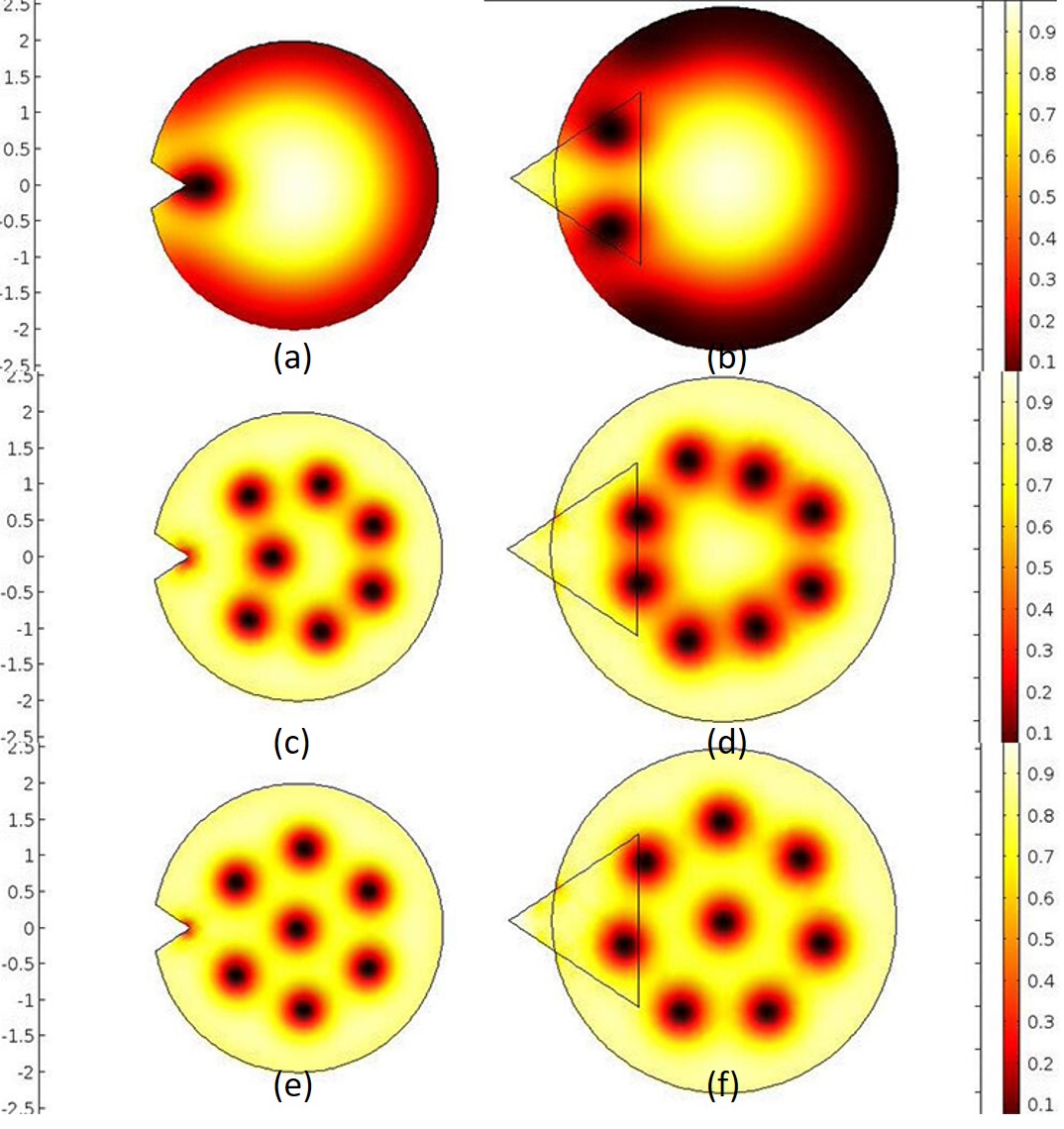}
\caption{Comparison of vortex dynamics in the samples with two types of geometrical defects.}
\label{defect}
\end{figure}
As illustrated in the Fig.~\ref{defect}, we find that it is easier for vortices to form at the defects with the obtuse angle, while the defects with acute angle tend to remain Meissner state. This phenomenon can be explained by the repulsive force between two vortices. For the acute angle situation, the vortices forming from two edges will repel each other, and the total force has non-zero component along the bisector of the acute angle, so the vortices tend to avoid the defects with acute angle.\par
Meanwhile, the vortices tend to be attracted by the surface for the Bean-Livingston image force\cite{deo1999hysteresis}, as also observed in the experiment\cite{geim2000non}. Due to the boundary condition of the superconductor, there will be an image anti-vortex at the opposite position of the interior vortex. And the interior vortex will be attracted by the image anti-vortex, but the interaction is short range, the vortex will only be attracted to the boundary when it is very close to the boundary, and will not feel the boundary when the distance from the boundary is larger than the magnetic field penetration depth $\lambda$.

\subsection{Vortex Pinning}
When the superconductors have at least one of defects, disorders and impurities, the magnetic field will get into the sample in form of vortices and pinned. By varying the parameters $\kappa, \alpha, \beta$ and changing the geometry of the samples, we can simulate different pinning states and investigate their properties\cite{koshelev2016optimization}. \par
\begin{figure}[H]
\centering
\includegraphics[width=0.4\textwidth]{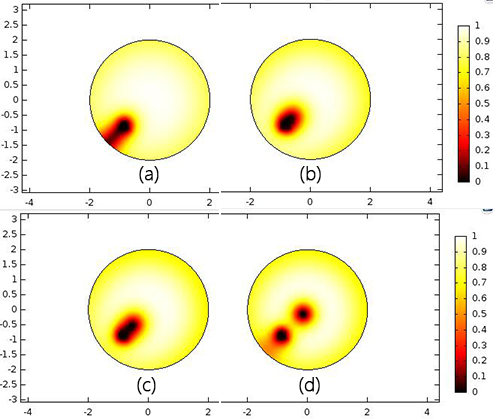}
\caption{When increasing the applied magnetic field, the vortex interacts with the defect.}
\label{pin}
\end{figure}

One model to describe the $\delta T$ pinning is achieved by modifying $\alpha$, namely the coefficient of the $\psi$ term in dimensionless TDGL, to be dependent on temperature.
\begin{equation}\label{pinmodel}
\begin{array}{l}
\left( {\frac{\partial }{{\partial t}} + i\kappa \Phi } \right)\Psi  = \\
 - {\left( {\frac{i}{\kappa }\nabla  + {\bf{A}}} \right)^2}\Psi  + \epsilon\left( {\bf{r}} \right)\left( {1 - t} \right)\Psi  - {\left| \Psi  \right|^2}\Psi
\end{array}
\end{equation}
where $t=T/T_c$ is reduced temperature, and $\epsilon(\mathbf{r})=1,0$ models the pinning center. As Fig.~\ref{pin} illustrated, vortex will be attracted by the defects which are often approximately modeled by an attractive parabolic wells\cite{reichhardt1996vortex}.\par
The defect obviously breaks the symmetry of the disk, and due to its attraction, the magnetic field will form a vortex from the boundary at the position which is the closet to the defect. And the vortex will site at the defect, until the magnetic field increased higher than the pinning attraction potential.
\section{Flux Pump}\label{fp}
High temperature superconductors are widely used to carry large current. However, the movable vortices in type II superconductors may cause large dissipation effects, in a result, reducing the critical current. The vortex dynamics deserve attention, because most of the HTSC are type II superconductors, and when the applied magnetic field is higher than $H_{c1}$, the superconductors will change into mixed state, in which movable vortices appears. In this section, we report using periodical magnetic field to enhance the supercurrent in a type-II superconducting wire with ratchet defects, which may correspond to vortex-rectification effect\cite{van2005vortex}.
\subsection{Geometry and Dynamics of the Ratchet Superconducting Wire}
\begin{figure}[H]
\centering
\includegraphics[width=0.35\textwidth]{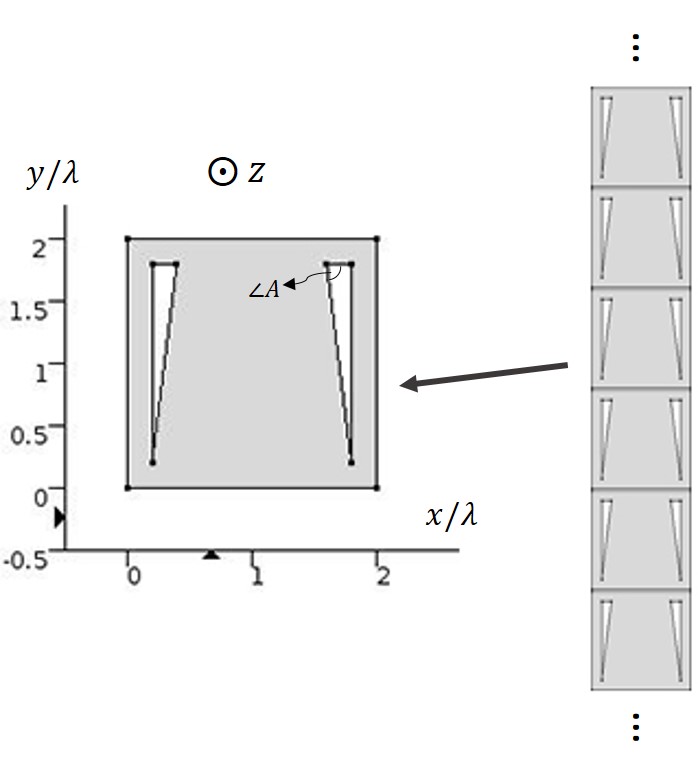}
\caption{Geometry of superconducting wire.}
\label{geowire}
\end{figure}

We designed the superconducting wire with strong unidirectional properties. As Fig.~\ref{geowire} illustrates, the infinite long superconducting wire is built by the blocks with the size of $2\lambda\times 2\lambda$, the top and bottom of each block are applied the periodical boundary condition. The geometrical defects and the two sides of the blocks are applied the boundary conditions of TDGL, as Eq.~\ref{nbc} suggests.\par
\begin{figure}[H]
\centering
\includegraphics[width=0.4\textwidth]{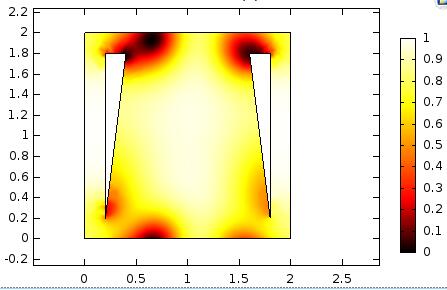}
\caption{Geometry of superconducting wire.}
\label{formavortex}
\end{figure}

We argue this geometry could easily make the vortices form and move in an efficient way and can be simply generalized to other geometry. As discussed in Section.~\ref{defecteffect}, we find that the defects with obtuse angle could easily form vortices, so the vortex will easily form at the angle $\angle A$(Fig.~\ref{formavortex}) when we turn on the magnetic field $B_0$ along $z$ direction. For the size of each block is comparable with the size of vortex, there will only form one vortex in each block.\par
\begin{figure}[H]
\centering
\includegraphics[width=0.4\textwidth]{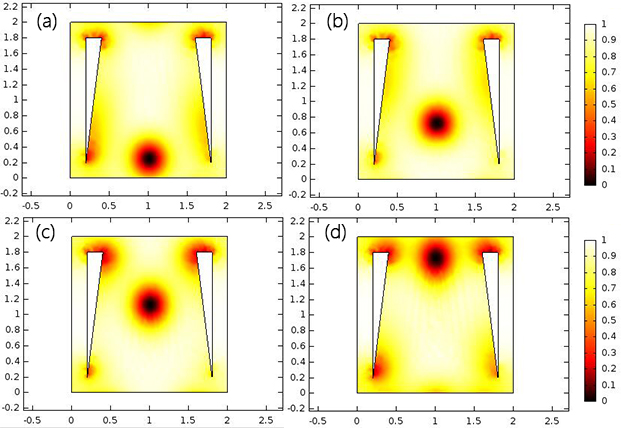}
\caption{Motion of vortex in the superconducting wire with extra periodical magnetic field, parameters are $B_0=1.3, B_c=0.8, \omega=0.12$.(a) $t=57$, (b) $t=70$, (c) $t=85$, (d) $t=100$.}
\label{motionvortex}
\end{figure}

To make the vortices move continuously, we add a periodical magnetic field along $z$ direction with wave like form, $\mathbf{B_p}=B_c \sin(\frac{2\pi}{L}y-\omega t)\mathbf{k}$, where $B_c, \omega$ is the magnitude and frequency of the periodical magnetic field, $L$ is the length along $y$ direction of each block. The total applied magnetic field is $\mathbf{B_a}=B_0\mathbf{k}+B_p\mathbf{k}$. Fig.~\ref{motionvortex} shows how the vortex move in the superconducting wire.

\subsection{Vortex Transportation and Supercurrent}
\begin{figure}[H]
\centering
\includegraphics[width=0.45\textwidth]{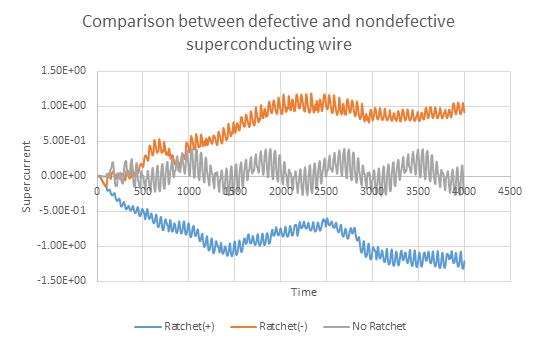}
\caption{Supercurrent of wire with ratchet and without ratchet, the $+,-$ are the ratchet direction, the ratchet in Fig.~\ref{geowire} is $+$. The blue, red and grey line are Ratchet(+), Ratchet(-) and No Ratchet respectively.}
\label{compwire}
\end{figure}

By integrating the current density along the top of each block, we can find the supercurrent in the superconducting wire,
\begin{equation}\label{supercurrentd}
{J_s} = \frac{1}{{2i\kappa }}({\Psi ^*}\nabla \Psi  - \Psi \nabla {\Psi ^*}) - {\left| \Psi  \right|^2}{\bf{A}},
\end{equation}
\begin{equation}\label{supercurrent}
{I_s} = \int_{top} J_s dl.
\end{equation}
By comparing with supercurrent in the superconducing wire without the ratchet-like defects, we find that the ratchet-like defects do enhance the unidirectional properties of the superconducting wire(Fig.~\ref{compwire}), the average supercurrent of the wire without ratchet-like defects is almost zero, but of the one with defects has non-zero value. The corresponding phenomenon of the superconducting wire without the ratchet-like defects is that the vortex didn't move from one block to another, the vortex only vibrates along with the periodical magnetic field $B_p$.

\begin{figure}[H]
\centering
\includegraphics[width=0.45\textwidth]{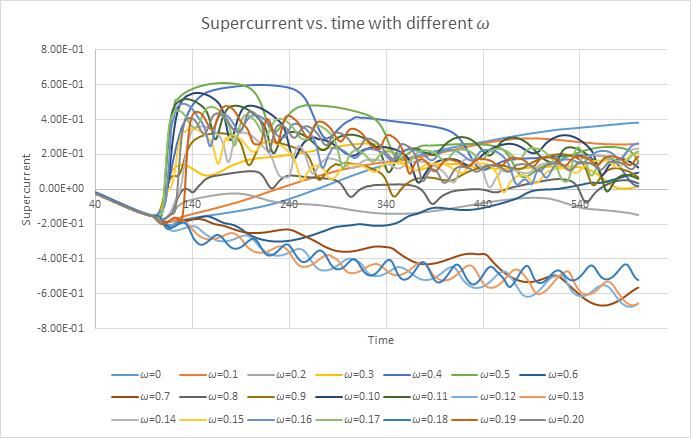}
\caption{$\omega \approx 0.12$ is the optimal frequency for this superconducting wire.}
\label{optcurrent}
\end{figure}

By tuning the frequency of $B_p$ to match the motion of vortex, we can make the vortex move in the highest speed, correspondingly, the current in the superconducting wire becomes largest. By using the parametric sweep, we can find the optimal frequency of $B_p$(Fig.~\ref{optcurrent})

\subsection{Analytical Analysis of the Optimal Frequency}
To find the optimal frequency of $B_p$, we turn to investigate the dynamics of single vortex. The overdamped equation for the single vortex is,
\begin{equation}\label{vortexdynamics}
\eta \mathbf{v}=\mathbf{f}_B+\mathbf{f}_{vv}
\end{equation}
where $\eta\approx B H_{c2}/c^2\rho_n$ ($\rho_n$ is the resistivity of normal-state) is the damping coefficient due to the dissipation processes inside and around the vortex cores\cite{bardeen1965theory}, $\mathbf{v}$ is the velocity of the vortex, $\mathbf{f}_B$ is the force exerted by the magnetic field, and $\mathbf{f}_{vv}$ is the force between two vortices. But for the periodical boundary condition, the vortex-vortex force can be neglected. \par
The periodical magnetic field will make the vortices move along the hypotenuse of the ratchets when the magnitude $B_c$ is below the Bean-Livingston surface barrier $H_s$. The force $\mathbf{f}_B$ is actually the type of vortex-vortex force\cite{reichhardt1996vortex},
\begin{equation}\label{magforce}
{{\bf{f}}_B} = \frac{1}{{2\pi {\mu _0}}}\frac{{{B_c}^2}}{{{\lambda ^3}}}\sum\limits_n {{K_1}(\frac{{\left| {{\bf{r}} - {{\bf{r}}_{B,n}}} \right|}}{\lambda })\frac{{{\bf{r}} - {{\bf{r}}_{B,n}}}}{{\left| {{\bf{r}} - {{\bf{r}}_{B,n}}} \right|}}}
\end{equation}
where ${\mathbf{r}}_{B,n}=(\omega Lt/2\pi  + nL){\bf{j}} + \omega L\tan \theta t/2\pi {\bf{i}}, n\in \mathbb{Z}$ is the position vector of the n\textsuperscript{th} vortex which is generated by the periodical magnetic field $B_p$, $\mathbf{r}$ is the position vector of the moving vortex. As shown in Fig.~\ref{k1x} and discussed before, the interaction decays exponentially in long range but ${{\rm{K}}_1}\left( {\frac{r}{\lambda }} \right) \approx \frac{\lambda }{r}$ at small $r$. Eq.~\ref{magforce} can be approximated as,
\begin{equation}\label{apmagforce}
{{\bf{f}}_B} = \frac{1}{{2\pi {\mu _0}}}\frac{{{B_c}^2}}{{{\lambda ^2}}}\sum\limits_{n \in nearest} {\frac{{{\bf{r}} - {{\bf{r}}_{B,n}}}}{{{{\left| {{\bf{r}} - {{\bf{r}}_{B,n}}} \right|}^2}}}}
\end{equation}
Eq.~\ref{vortexdynamics} can be simplified as,
\begin{equation}\label{simpleeq}
x'\left( t \right) = \frac{s}{{ - kt + x\left( t \right)}}
\end{equation}
where $k=\omega L/2\pi$ and $s=\frac{1}{{2\pi {\mu _0}\eta }}\frac{{{B_c}^2}}{{{\lambda ^2}}}$. By solving Eq.~\ref{simpleeq}, we can find the average velocity is,
\begin{equation}\label{velocity}
\left\langle v \right\rangle  \simeq \frac{{{{\rm{e}}^{\frac{{k\left( {l - 2{x_0}} \right)}}{s}}}{k^2}\left( {l - {x_0}} \right)}}{{s - k{x_0} - {{\rm{e}}^{\frac{{k\left( {l - 2{x_0}} \right)}}{s}}}\left( {s + k\left( { - l + {x_0}} \right)} \right)}}
\end{equation}
where $x_0$ is the initial position of the moving vortex. The supercurrent density can be characterized by the average velocity according to,
\begin{equation}\label{supercurrentt}
\mathbf{j} = \frac{1}{{2i}}({\Psi ^*}(\frac{\mathbf{p}}{m}\Psi ) + {(\frac{\mathbf{p}}{m}\Psi )^*}\Psi ) = {\mathop{\rm Re}\nolimits} ({\Psi ^*}(\frac{\mathbf{p}}{m}\Psi ))\sim \rho\mathbf{v}
\end{equation}
And when $k$ satisfies $ k\sim \frac{2s}{L-x_0}$, the velocity reaches the maximum.

\section{Summary}\label{sum}
In this study, time-dependent Ginzburg-Landau equation is solved in different geometrical regions numerically by finite element method. The distribution of the order parameter is vividly depicted and the magnetic vortices can be found. Some derived quantities, such as magnetization and current density, are also calculated. By modifying the parameters in the TDGL equation, the pinning effect is investigated. In the last part, we investigate a special superconducting wire with ratchet-like defects, in which the unidirectional motion of the vortices can enhance the supercurrent.\par
However, some limitations in this study need to be fixed in the future. The TDGL equations(Eq.~\ref{tdgl1}, Eq.~\ref{tdgl2}) used in this study are only suitable for the s-wave superconductors such as niobium, but most of the type II superconductors, such as BSCCO and YBCO, are d-wave or even more complex pairing symmetry, thus, Eq.~\ref{tdgl1}, Eq.~\ref{tdgl2} are no longer suitable. Some researchers have generalized the Ginzburg-Landau equation for d-wave and p-wave superconductors\cite{xu1996structures}\cite{zhu1997ginzburg}, these can be adapted for our future study.

\nocite{*}

\bibliography{ws-rv-sample}

\end{document}